\documentclass[10pt,prb,twocolumn,showpacs,reprint]{revtex4-1}
\usepackage{graphicx,amsmath,amssymb}
\usepackage{epsfig}
\usepackage[usenames,dvipsnames]{color}
\usepackage[colorlinks=true,citecolor=blue,linkcolor=blue]{hyperref}
\begin{document}

\renewcommand{\Re}{\mathop{\mathrm{Re}}}
\renewcommand{\Im}{\mathop{\mathrm{Im}}}
\renewcommand{\b}[1]{\mathbf{#1}}
\renewcommand{\u}{\uparrow}
\renewcommand{\d}{\downarrow}
\newcommand{\bsigma}{\boldsymbol{\sigma}}
\newcommand{\blambda}{\boldsymbol{\lambda}}
\newcommand{\Tr}{\mathop{\mathrm{Tr}}}
\newcommand{\sgn}{\mathop{\mathrm{sgn}}}
\newcommand{\sech}{\mathop{\mathrm{sech}}}
\newcommand{\diag}{\mathop{\mathrm{diag}}}
\newcommand{\half}{{\textstyle\frac{1}{2}}}
\newcommand{\sh}{{\textstyle{\frac{1}{2}}}}
\newcommand{\ish}{{\textstyle{\frac{i}{2}}}}
\newcommand{\thf}{{\textstyle{\frac{3}{2}}}}
\newcommand{\be}{\begin{equation}}
\newcommand{\ee}{\end{equation}}

\title{Topological order in a correlated Chern insulator}

\author{Joseph Maciejko$^1$ and Andreas R\"{u}egg$^2$}

\affiliation{$^1$Princeton Center for Theoretical Science, Princeton University, Princeton, New Jersey 08544, USA\\
$^2$Department of Physics, University of California, Berkeley, California 94720, USA}

\date\today

\begin{abstract}
We study the effect of electron-electron interactions in a spinful Chern insulator. For weak on-site repulsive interactions at half-filling, the system is a weakly correlated Chern insulator adiabatically connected to the noninteracting ground state, while in the limit of infinitely strong repulsion the system is described by an effective spin model recently predicted to exhibit a chiral spin liquid ground state. In the regime of large but finite repulsion, we find an exotic gapped phase with characteristics partaking of both the noninteracting Chern insulator and the chiral spin liquid. This phase has an integer quantized Hall conductivity $2e^2/h$ and quasiparticles with electric charges that are integer multiples of the electron charge $e$, but the ground state on the torus is four-fold degenerate and quasiparticles have fractional statistics. We discuss how these unusual properties affect the outcome of a charge pumping experiment and, by deriving the topological field theory, elucidate that the topological order is of the exotic $\mathbb{Z}_2$ double-semion type.
\end{abstract}


\maketitle

Over the past few years, the study of topological insulators\cite{hasan2010,qi2011} has become one of the most active areas of condensed matter research. The prototypical topological insulator is the Chern insulator (CI), first theoretically proposed by Haldane,\cite{haldane1988} which exhibits the integer quantum Hall effect (QHE) in the absence of any external magnetic fields. Many concrete proposals to realize this new state of matter have followed Haldane's original idea, culminating in the recent theoretical prediction\cite{yu2010} and experimental discovery\cite{chang2013} of the CI in thin films of Cr-doped (Bi,Sb)$_2$Te$_3$. Although electron-electron interactions are needed to establish ferromagnetic order in these films and break time-reversal symmetry as required for the existence of a QHE, this effect is described by mean-field theory and the resulting CI state is weakly correlated. A burning question in the field of topological insulators is whether strong electron correlation effects can give rise to qualitative changes in their properties.\cite{[{For a recent review, see }]hohenadler2013}

\begin{figure}
\includegraphics[width=1\linewidth]{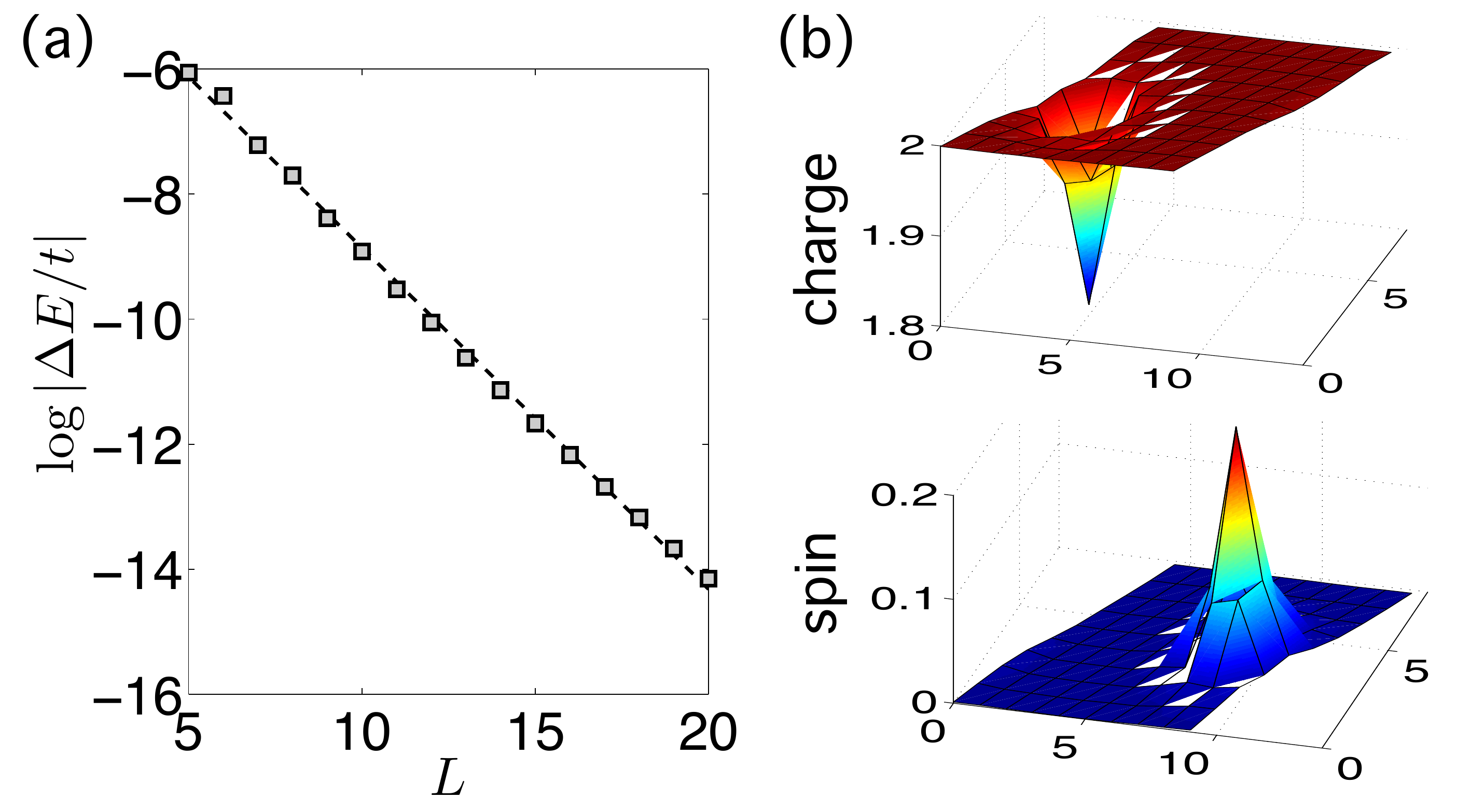}
\caption{(Color online.) (a) Topological ground state degeneracy in the CI*: energy difference between two different global $\mathbb{Z}_2$ flux configurations on finite tori with $L\times L$ unit cells. The dashed curve is a fit corresponding to an exponential dependence on $L$. (b) Spin-charge separation in the CI*: a single hole splits into two anyons, a charge $-e$ particle without spin (quasiparticle $l_1$ in Table~\ref{TableI}) and a neutral particle with spin 1/2 (bound state of quasiparticle $l_1$ and $l_4$). Results are obtained from self-consistent calculations on the honeycomb lattice model with Hubbard $U=26t$ and $t_2=it$.}
\label{fig:TO}
\end{figure}

Similar to the spin-polarized $\nu=1$ integer QH state, the simplest model of CI such as Haldane's original model describes noninteracting spinless electrons and has a Hall conductivity $\sigma_{xy}=Ce^2/h$ where the Chern number $C=\pm 1$. The simplest type of interaction one can add is a nearest-neighbor density-density interaction. Many recent studies have focused on partially filled bands, which correspond to a metal in the noninteracting limit and lead at certain fillings to a fractional CI~\cite{tang2011,sun2011,neupert2011,sheng2011,regnault2011} for strong enough interactions and sufficiently flat bands. In this paper we consider a completely filled valence band (half-filling), corresponding to a CI in the noninteracting limit. Analytical and numerical studies~\cite{cai2008,wang2010,varney2010, *varney2011} of the spinless interacting problem show that the CI persists up to a critical interaction strength beyond which the ground state develops charge density wave order. Once the electron spin is considered, two possibilities arise. The first is to combine two copies of the CI with $C_\uparrow=1$ for spin up and $C_\downarrow=-1$ for spin down. The resulting noninteracting model is the time-reversal invariant quantum spin Hall (QSH) insulator~\cite{[{For a recent review, see, e.g., }]maciejko2011} in the limit of conserved $z$ component of spin.\cite{kane2005,bernevig2006} In this case, the simplest type of interaction to consider is an on-site Hubbard interaction, and the resulting interacting problem at half-filling has been the focus of intense recent study.\cite{hohenadler2013} The second possibility is to consider two copies of the CI with the same Chern number $C_\uparrow=C_\downarrow=\pm 1$ for both spins. The resulting noninteracting model at half-filling is a CI with total Hall conductivity $\sigma_{xy}=\pm 2e^2/h$. In this paper we consider the effect of an on-site Hubbard interaction $U$ in such a system. This problem, which has received considerably less attention than the interacting QSH insulator, has been studied recently using the slave-rotor technique~\cite{he2011,he2012} and the mapping to a spin model.\cite{nielsen2013} In both cases a ground state with topological order corresponding to the Kalmeyer-Laughlin chiral spin liquid state~\cite{kalmeyer1987,*kalmeyer1989,schroeter2007,thomale2009} was found in the phase diagram. This result corresponds to the $U\rightarrow\infty$ limit where charge degrees of freedom are frozen and spin-charge separation occurs. Here, motivated by recent work on the interacting QSH insulator~\cite{Ran:2008,ruegg2012} we are interested in the possibility of new phases occurring at intermediate $U$. Besides the weakly correlated CI at small $U$ and the chiral spin liquid at large $U$, we find within the $\mathbb{Z}_2$ slave-spin representation~\cite{huber2009,ruegg2010,nandkishore2012} a new chiral topological phase at intermediate $U$ which we dub the CI* phase by analogy with the QSH* phase in Ref.~\onlinecite{ruegg2012}. Similar to the QSH* phase, the CI* phase is characterized by an emergent deconfined $\mathbb{Z}_2$ gauge field, a four-fold topological ground state degeneracy on the torus, and spin-charge separated excitations with fractional statistics (Fig.~\ref{fig:TO} and Table~\ref{TableI}). But in contrast to the QSH* phase, the topological field theory of the CI* phase reveals a topological order of the $\mathbb{Z}_2$ double-semion type supplemented by chiral fermions.

We consider a class of 2D electron lattice models of the form
\begin{align}\label{H}
H=\sum_{rr'}\sum_\alpha t_{rr'}c_{r\alpha}^\dag c_{r'\alpha}+U\sum_r\left(\sum_\alpha n_{r\alpha}-1\right)^2,
\end{align}
where $c_{r\alpha}^\dag$ ($c_{r\alpha}$) is a creation (annihilation) operator for an electron of spin $\alpha=\uparrow,\downarrow$ at site $r$, $n_{r\alpha}=c_{r\alpha}^\dag c_{r\alpha}$ is the number of electrons of spin $\alpha$ on site $r$, $t_{rr'}=t_{r'r}^*$ is a spin-independent hopping amplitude and $U$ is the on-site Hubbard interaction. We consider the half-filled case $\langle\sum_\alpha n_{r\alpha}\rangle=1$, and choose the hopping amplitude $t_{rr'}$ to be that of a spinless two-band CI with (say) Chern number $1$ in the valence band. Although we will present numerical results for Haldane's honeycomb lattice model~\cite{haldane1988} later on in the paper, we begin with a more general discussion of the possible phases and their expected universal properties which are independent of the details of the lattice model.

The model (\ref{H}) can be first investigated in two extreme limits. In the limit $U=0$, the system is a noninteracting CI with total Chern number $C=2$ and Hall conductivity $\sigma_{xy}=2e^2/h$. Because this is a gapped state, we expect $\sigma_{xy}$ to remain quantized for nonzero but small $U$, with the ground state evolving adiabatically from a noninteracting CI to a weakly correlated CI. In the limit $U\rightarrow\infty$, one can derive an effective low-energy Hamiltonian in the subspace of one electron per site which, owing to the full $SU(2)$ spin rotation symmetry of the Hamiltonian (\ref{H}), is an $SU(2)$ invariant $S=1/2$ spin model. For a particular choice of $t_{rr'}$ on the square lattice, such a spin model was recently studied numerically~\cite{nielsen2013} and argued to have a chiral spin liquid ground state.

In order to study the possible phases of (\ref{H}) at intermediate values of $U$, we make use of the recently introduced $\mathbb{Z}_2$ slave-spin theory for correlated electron systems.\cite{huber2009,ruegg2010,nandkishore2012,ruegg2012} This theory is based on the simple observation that the Hubbard interaction energy in (\ref{H}) depends only on the total occupation of site $r$ \emph{modulo 2}, which can be represented by an Ising variable $\tau^z_r$, viz. $\tau^z_r=1$ for unoccupied or doubly occupied sites and $\tau^z_r=-1$ for singly occupied sites. The electron operator $c_{r\alpha}^{(\dag)}$ is written as a product of a slave-fermion operator $f_{r\alpha}^{(\dag)}$ and an Ising slave-spin $\tau^x_r$,
\begin{align}\label{parton}
c_{r\alpha}^{(\dag)}=f_{r\alpha}^{(\dag)}\tau^x_r,
\end{align}
where $\tau^x_r$ and $\tau^z_r$ obey the (anti)commutation relations of the associated Pauli matrices, so that $\tau^x_r$ flips the sign of $\tau^z_r$ since creating/annihilating an electron changes the occupation modulo 2. The Hamiltonian (\ref{H}) is then written in terms of the slave-fermions and slave-spins as
\begin{align}\label{Hss}
H=&\sum_{rr'}\sum_\alpha t_{rr'}\tau^x_r\tau^x_{r'}
f_{r\alpha}^\dag f_{r'\alpha}
+\frac{U}{2}\sum_r\left(\tau^z_r+1\right).
\end{align}
While the original Hamiltonian (\ref{H}) acts in the Hilbert space of physical electron states $\{|\psi_\textrm{phys}\rangle\}$, the Hamiltonian (\ref{Hss}) acts in an enlarged Hilbert space which contains unphysical states $\{|\psi_\textrm{unphys}\rangle\}$ such as states with a single slave-fermion $\sum_\alpha f_{r\alpha}^\dag f_{r\alpha}=1$ but formally zero/double occupancy $\tau^z_r=1$. When evaluating the partition function, one should sum only over physical states. This is accomplished by introducing a projector $P$ inside the partition sum $Z=\Tr\left(e^{-\beta H}P\right)$ where $P$ is defined such that $P|\psi_\textrm{phys}\rangle=|\psi_\textrm{phys}\rangle$ and $P|\psi_\textrm{unphys}\rangle=0$. The four-operator term $\tau^x\tau^xf^\dag f$ in Eq.~(\ref{Hss}) can be decoupled with auxiliary fields. After performing a saddle-point approximation with respect to these auxiliary fields and treating fluctuations within the constant-amplitude approximation,\cite{supplement} which corresponds to first-order mean-field theory in the sense of Wen,\cite{WenBook} $Z$ becomes the partition function of a $\mathbb{Z}_2$ gauge theory~\cite{kogut1979} in 3D Euclidean spacetime with bosonic and fermionic matter in the fundamental representation,
\begin{align}\label{Zgauge3D}
Z=\int D\bar{f}_{i\alpha}Df_{i\alpha}\sum_{\{\tau^x_i\}}
\sum_{\{\sigma_{ij}\}}e^{-S[\bar{f},f,\tau^x,\sigma]},
\end{align}
with an action $S=S_{\tau^x}+S_f+S_B$ such that
\begin{align}
S_{\tau^x}&=-\kappa\sum_{ij}\tau^x_i\sigma_{ij}\tau^x_j,\nonumber\\
S_f&=-\sum_{ij}\sum_\alpha t_{ij}\bar{f}_{i\alpha}\sigma_{ij}f_{j\alpha},\nonumber\\
e^{-S_B}&=\prod_{i,j=i-\hat{\tau}}\sigma_{ij},\nonumber
\end{align}
where $\sigma_{ij}=\pm 1$ is a spacetime $\mathbb{Z}_2$ gauge field, $i,j$ are sites on a 3D spacetime lattice, and $t_{ij}$ is proportional to $t_{rr'}$ on spatial links and equal to $-1$ on temporal links. $S_B$ is a Berry phase term~\cite{senthil2000} which corresponds to a background static $\mathbb{Z}_2$ charge on every site.\cite{moessner2001c} The constant $\kappa$ depends on the Hubbard interaction strength $U$,
\begin{align}\label{kappa}
\kappa=\frac{1}{2}\ln\coth\left(\frac{\epsilon U}{2}\right),
\end{align}
where $\epsilon$ is a short-time cutoff.\cite{epsilon}

We now discuss the possible phases of the gauge theory (\ref{Zgauge3D}). In the noninteracting limit $U=0$, Eq.~(\ref{kappa}) implies that $\kappa\rightarrow\infty$. In this limit, the gauge fields are frozen~\cite{fradkin1979} and can be gauged away, e.g., $\sigma_{ij}=1$. The only configurations which contribute to the partition function in the $\kappa\rightarrow\infty$ limit are those with the slave-spins ferromagnetically ordered, e.g., $\tau^x_i=1$ on every site (or gauge equivalent configurations). Equation~(\ref{parton}) then implies that the electron and slave-fermion operators are proportional, and we recover the noninteracting CI. In the $U\rightarrow\infty$ limit, Eq.~(\ref{kappa}) implies that $\kappa\rightarrow 0$. In this limit the slave-spins can be integrated out. Tracing over $\sigma_{ij}$ on spatial links generates interactions between slave-fermions,
\begin{align}
\delta S_f^\textrm{eff}=-\frac{1}{2}\sum_{rr'}\sum_\tau\sum_{\alpha\beta}
(t_{rr'})^2\bar{f}_{r\alpha\tau}f_{r'\alpha\tau}
\bar{f}_{r\beta\tau}f_{r'\beta\tau}
+\ldots,\nonumber
\end{align}
where $\ldots$ involves higher multiples of four slave-fermion operators, while tracing over $\sigma_{ij}$ on temporal links imposes the local constraint $\sum_\alpha f_{r\alpha}^\dag f_{r\alpha}=1$ on every site $r$. Therefore the $U\rightarrow\infty$ limit corresponds as expected to a $S=1/2$ spin model, which can exhibit a variety of ground states depending on the form of $t_{rr'}$.  Ref.~\onlinecite{nielsen2013} argues that a special choice of $t_{rr'}$ on the square lattice can give rise to a chiral spin liquid.

We now discuss the regime of large but finite $U$. In general, using the Poisson summation formula we can rewrite a $\mathbb{Z}_p$ gauge theory (here $p=2$) as a compact $U(1)$ gauge theory coupled to a charge-$p$ integer-valued link variable $n_{ij}$ with no dynamics,\cite{ukawa1980}
\begin{align}
Z=\int D\bar{f}_{i\alpha}Df_{i\alpha}Da_{ij}\sum_{\{\tau^x_i\}}
\sum_{\{n_{ij}\}} e^{-S[\bar{f},f,\tau^x,\sigma_{ij}=e^{ia_{ij}},n]},\nonumber
\end{align}
where $S=S_{\tau^x}+S_f+S_n+S_B$ with
\begin{align}
S_{\tau^x}&=-\kappa\sum_{ij}\tau^x_ie^{ia_{ij}}\tau^x_j,
\nonumber\\
S_f&=-\sum_{ij}\sum_\alpha t_{ij}\bar{f}_{i\alpha}e^{i(a_{ij}+eA_{ij})}
f_{j\alpha},\nonumber\\
S_n&=-ip\sum_{ij}n_{ij}a_{ij},\nonumber\\
e^{-S_B}&=\prod_{i,j=i-\hat{\tau}}e^{ia_{ij}},\nonumber
\end{align}
where $a_{ij}$ is a compact $U(1)$ gauge field and we have introduced the external $U(1)$ electromagnetic gauge potential $A_{ij}$ which couples to the electrons with charge $e$. The summation over $n_{ij}$ has the effect of discretizing $a_{ij}$ in integer multiples of $2\pi/p$. For large but finite $U$, $\kappa$ is small. The slave-spins are gapped and can be integrated out perturbatively,\cite{fradkin1979} resulting in a lattice version of the Maxwell term for $a_{ij}$. Because the band structure of the slave-fermions is gapped, we expect on general grounds~\cite{kogut1979} that the resulting gauge theory admits a gapped deconfined phase for large but finite values of $U$. In the rest of the paper we focus on the physics of this phase which we denote the CI* phase.

\begin{table}
\begin{tabular}{c||c|c|c|c}
quasiparticle & $Q$ & $S_z$ & $\theta_{ll}$ & $\theta_{ll'}$ \\
\hline
$l_1=(1,0,0,0)$ & $-e$ & $0$ & $\pi/2$ & $\theta_{13}=\pi$ \\
$l_2=(0,-1,0,0)$ & $e$ & $0$ & $-\pi/2$ & $\theta_{23}=-\pi$ \\
$l_3=(1,-1,0,0)$ & $0$ & $0$ & $0$ & $\theta_{31}=-\theta_{32}=\pi$ \\
$l_4=(0,0,1,0)$ & $e$ & $1/2$ & $\pi$ &  \\
$l_5=(0,0,0,1)$ & $e$ & $-1/2$ & $\pi$ & 
 \end{tabular}
\caption{Electromagnetic charge $Q$, spin $S_z$, self-statistics $\theta_{ll}$ and nontrivial mutual statistics $\theta_{ll'}$ of quasiparticles in the CI* phase, where $l_i$ is the gauge charge vector~\cite{wen1995} of quasiparticle $i$.}
\label{TableI}
\end{table}

In the CI* phase, the gauge field $a_{ij}$ is essentially free and we can take the continuum limit $a_{ij}\rightarrow a_\mu$. Following Senthil and Fisher,\cite{senthil2000} we expect that the Berry phase term $S_B$ can affect the position of the phase boundary for the deconfined phase but not the universal properties of this phase. On the other hand, $S_n$ is essential to remember the $\mathbb{Z}_2$ nature of the gauge theory in the continuum $U(1)$ description. The link variable $n_{ij}\rightarrow n_\mu$ obeys the constraint $\partial_\mu n_\mu=0$ to preserve the gauge invariance of the action under a $U(1)$ gauge transformation of $a_\mu$. Solving the constraint by $n_\mu=\frac{1}{2\pi}\epsilon_{\mu\nu\lambda}\partial_\nu b_\lambda$ where $b_\mu$ is a compact $U(1)$ gauge field and passing to a real time description, we obtain
\begin{align}
S_n\rightarrow\frac{p}{2\pi}\int d^3x\,\epsilon^{\mu\nu\lambda}b_\mu\partial_\nu a_\lambda,\nonumber
\end{align}
i.e., the (2+1)D level-$p$ $BF$ term.\cite{blau1991,hansson2004} Finally, we consider the slave-fermion action $S_f$. In a hydrodynamic approach,\cite{wen1995} we introduce a conserved $U(1)$ current $j_\sigma^\mu=\frac{1}{2\pi}\epsilon^{\mu\nu\lambda}\partial_\nu a^\sigma_\lambda$ for slave-fermions of each spin $\sigma=\uparrow,\downarrow$,\cite{spin}
where $a_\mu^\uparrow,a_\mu^\downarrow$ are compact $U(1)$ gauge fields. Considering that slave-fermions of each spin form a CI with Chern number $C_\uparrow=C_\downarrow=1$, we obtain the effective Lagrangian of the CI* phase,~\cite{supplement}
\begin{align}\label{LCI}
\mathcal{L}_\textrm{CI*}&=\frac{1}{\pi}\epsilon^{\mu\nu\lambda}b_\mu\partial_\nu a_\lambda
-\frac{e}{2\pi}\epsilon^{\mu\nu\lambda}A_\mu\partial_\nu (a_\lambda^\uparrow+a_\lambda^\downarrow)\nonumber\\
&+\sum_{\sigma=\uparrow,\downarrow}\left(
\frac{1}{4\pi}\epsilon^{\mu\nu\lambda}a_\mu^\sigma\partial_\nu a_\lambda^\sigma+\frac{1}{2\pi}\epsilon^{\mu\nu\lambda}a_\mu^\sigma\partial_\nu a_\lambda\right),
\end{align}
which is the main result of this work. From Eq.~(\ref{LCI}) we can extract all the universal topological properties of the CI* phase, such as the $K$-matrix,\cite{wen1995,lu2012}
\begin{align}\label{Kmatrix}
K=\left(\begin{array}{cccc}
0 & 2 & 1 & 1\\
2 & 0 & 0 & 0\\
1 & 0 & 1 & 0\\
1 & 0 & 0 & 1
\end{array}\right),
\end{align}
in the basis $(a_\mu,b_\mu,a_\mu^\uparrow,a_\mu^\downarrow)$. The ground state degeneracy on the torus is $|\det K|=4$, as for phases with $\mathbb{Z}_2$ topological order.\cite{wen1991} Unlike such phases however, the CI* phase is a chiral topological phase with Hall conductivity
\begin{align}\label{HallConductivity}
\sigma_{xy}=\frac{e^2}{h}t^TK^{-1}t=\frac{2e^2}{h},
\end{align}
where $t=(0,0,1,1)$ is the electromagnetic charge vector. The $K$-matrix (\ref{Kmatrix}) can be made diagonal by a $GL(4,\mathbb{Z})$ transformation $W$,\cite{supplement}
\begin{align}
K&\rightarrow W^TKW=\diag(2,-2,1,1),\label{Kdiag}\\
t&\rightarrow W^Tt=(-2,2,1,1),\nonumber
\end{align}
which describes the direct sum of the double semion model~\cite{levin2005} and a CI with Chern number $C_\uparrow=C_\downarrow=1$. The quasiparticle excitations in the CI* phase and their quantum numbers can be read off from the $K$-matrix (Table~\ref{TableI}). The quasiparticles $l_1$ (semion), $l_2$ (antisemion), and $l_3$ (semion-antisemion bound state) have the same statistics as the excitations of the double semion model, but in contrast to their occurrence in spin models, here these excitations are electrically charged. Their charge and statistics can be understood as follows. An applied $\pi$ flux in a $\nu=1$ integer QH state traps a fermion mode with $Q=\pm e/2$, and the resulting flux-charge composite behaves as an anyon with statistical angle $\frac{\pi}{4}$.\cite{weeks2007} The semions/antisemions in our model likewise arise from in-gap modes of the slave-fermions with total Chern number 2, which are bound to dynamical $\mathbb{Z}_2$ vortices. The quasiparticles $l_4$ and $l_5$ are identified with the original electrons, and are topologically equivalent to the vacuum.\cite{kitaev2006} Although the quasiparticle spin $S_z$ in Table~\ref{TableI} is only meaningful in the presence of $U(1)$ spin rotation symmetry, the other properties of the CI* phase are robust under perturbations which break this symmetry but do not close the bulk gap.

\begin{figure}
\includegraphics[width=1\linewidth]{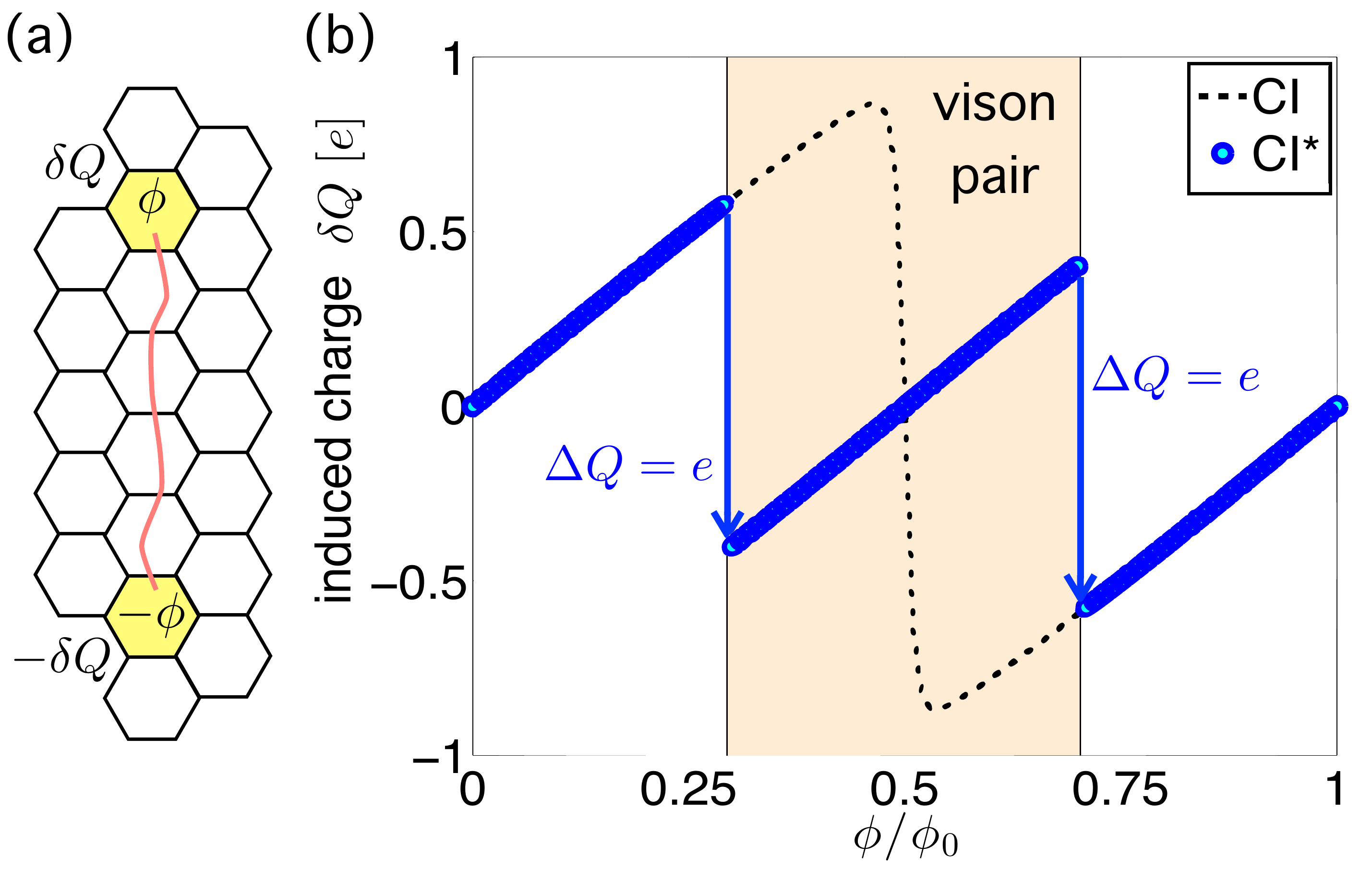}
\caption{(Color online). (a) Charge pumping between two hexagons threaded by external fluxes $\phi$ and $-\phi$ and separated by $D$ unit cells. (b) Transferred charge as a function of $\phi$ in the CI (dashed curve) and the CI* (blue circles), in units of the flux quantum $\phi_0=h/e$ for $D=8$. The shaded area denotes the parameter range where a spontaneously created $\mathbb{Z}_2$ vortex pair (semion/antisemion) partly screens the external fluxes.}
\label{fig:CP}
\end{figure}

The existence of protected edge excitations can also be inferred from the $K$-matrix.\cite{wen1995} $K$ has an imbalance of positive and negative eigenvalues [Eq.~(\ref{Kdiag})], which translates into an imbalance of right-moving and left-moving edge excitations on a manifold with boundary. Such a chiral edge cannot be completely gapped out by local perturbations.\cite{haldane1995} However, because the slave-spins are disordered in the CI* phase ($\langle\tau^x_i\rangle=0$), the electron and slave-fermion operators are not proportional. We therefore expect that the electron spectral function on the edge is gapped while correlation functions of on-site electron bilinears $c_{r\alpha}^\dag c_{r\beta}$ or $c_{r\alpha}c_{r\beta}$ are gapless, as in the QSH* phase.\cite{ruegg2012}

To further study the CI* phase we use the self-consistent mean-field framework in the Hamiltonian language.\cite{ruegg2010,ruegg2012} This further approximation, which corresponds to zeroth-order mean-field theory in the sense of Wen,\cite{WenBook} neglects the dynamical fluctuations of the $\mathbb{Z}_2$ gauge field and involves unphysical states $|\psi_\textrm{unphys}\rangle$. However, the CI* zeroth-order mean-field ground state is stable to $\mathbb{Z}_2$ gauge fluctuations because it is gapped.\cite{WenBook} In this scheme, the bond energies of the free slave-fermion model provide the exchange couplings $J_{rr'}=\sum_{\alpha}(t_{rr'}\langle f_{r\alpha}^{\dag}f_{r'\alpha}\rangle+{\rm c.c.})$ of the transverse-field Ising model, whereas the correlation functions $g_{rr'}=\langle\tau_r^x\tau_{r'}^x\rangle$ of the transverse-field Ising model renormalize the hopping parameters of the slave-fermions.\cite{supplement} Performing unrestricted mean-field calculations in the CI* phase on finite systems, we find a four-fold topological ground state degeneracy on the torus [Fig.~\ref{fig:TO}(a)] as well as spin-charge separated excitations above the ground state [Fig.~\ref{fig:TO}(b)]. For these numerical results, we studied the spinful Haldane-Hubbard model on the honeycomb lattice with $U=26t$ and a complex second-neighbor hopping $t_2=i t$ where $t$ is the nearest-neighbor hopping.

Although the Hall conductivity (\ref{HallConductivity}) of the CI* is the same as in the weakly interacting CI, interesting differences appear beyond the linear response regime (Fig.~\ref{fig:CP}). We study the induced charge in a charge-pumping setup where an external flux $\phi$ is threaded through one hexagon and a flux $-\phi$ through an other hexagon separated by 8 unit cells. While for small fluxes the pumped charge follows the expected relation $\delta Q=C\phi/\phi_0$ with $C=2$, we observe a sudden jump of $\Delta Q=e$ for $\phi$ slightly larger then $\phi_0/4$. This jump is associated with the spontaneous creation of a $\mathbb{Z}_2$ vortex pair (semion/antisemion) which is localized at the flux-pierced hexagons and partially screens the external fluxes.\cite{supplement} A second jump of $\Delta Q=e$ occurs at a flux slightly less than $3\phi_0/4$. This behavior is in sharp contrast to the charge pumping observed in the conventional CI (dashed curve) where the screening of the fluxes by $\mathbb{Z}_2$ vortices cannot occur.

In summary, we predict that a new chiral topological phase of matter, the CI*, can arise when a spinful CI is subjected to a strong on-site Hubbard interaction. To be clear, our result is not a rigorous proof that the CI* phase does occur in a specific lattice Hamiltonian of the type (\ref{H}). What we do show is that the CI* is a possible ground state of such Hamiltonians, and that if it is realized in a specific Hamiltonian, its topological field theory is Eq.~(\ref{LCI}). Some properties of the CI* are reminiscent of noninteracting topological phases (integer quantized Hall conductivity and integer charge of quasiparticles) while others indicate the presence of topological order (nonzero ground state degeneracy on the torus and fractional statistics of quasiparticles). To test these predictions, a determination of the phase diagram of Hamiltonian (\ref{H}) via exact numerical diagonalization or density-matrix renormalization group studies is desirable. The interesting question of whether the chiral spin liquid of Ref.~\onlinecite{nielsen2013} can be reached via a continuous transition from the CI* can perhaps be studied analytically~\cite{bais2009} using the topological field theory (\ref{LCI}) as a starting point, and is left for future work.

After the completion of this work, we learned of a related study of the CI* phase by Zhong \emph{et al.}~\cite{zhong2013}

We thank S.-P.~Kou, Y.-M.~Lu, R.~Nandkishore, T.~Senthil, R.~Thomale and A.~Vishwanath for insightful discussions. We acknowledge financial support from the Simons Foundation (JM) and the Swiss National Science Foundation (AR), as well as the hospitality of the Aspen Center for Physics (NSF Grant \#1066293) where this work was initiated.

\bibliography{qah}

\begin{thebibliography}{50}
\bibitem [{eps()}]{epsilon}%
  \BibitemOpen
  \href@noop {} {}\bibinfo {note} {For simplicity, we kept only the dominant
  nearest-neighbor coupling and chose the value of cutoff $\epsilon$ in a way
  to make the couplings isotropic in space and time}\BibitemShut {NoStop}%
\end{thebibliography}

\end{document}


\renewcommand{\Re}{\mathop{\mathrm{Re}}}
\renewcommand{\Im}{\mathop{\mathrm{Im}}}
\renewcommand{\b}[1]{\mathbf{#1}}
\renewcommand{\u}{\uparrow}
\renewcommand{\d}{\downarrow}
\newcommand{\bsigma}{\boldsymbol{\sigma}}
\newcommand{\blambda}{\boldsymbol{\lambda}}
\newcommand{\Tr}{\mathop{\mathrm{Tr}}}
\newcommand{\sgn}{\mathop{\mathrm{sgn}}}
\newcommand{\sech}{\mathop{\mathrm{sech}}}
\newcommand{\diag}{\mathop{\mathrm{diag}}}
\newcommand{\half}{{\textstyle\frac{1}{2}}}
\newcommand{\sh}{{\textstyle{\frac{1}{2}}}}
\newcommand{\ish}{{\textstyle{\frac{i}{2}}}}
\newcommand{\thf}{{\textstyle{\frac{3}{2}}}}
\newcommand{\be}{\begin{equation}}
\newcommand{\ee}{\end{equation}}

\renewcommand{\thetable}{S\Roman{table}}
\renewcommand{\thefigure}{S\arabic{figure}}
\renewcommand{\thesubsection}{S\arabic{subsection}}
\renewcommand{\theequation}{S\arabic{equation}}

\setcounter{secnumdepth}{1}
\setcounter{equation}{0}
\setcounter{figure}{0}
\setcounter{section}{0}

\title{Supplemental material for ``Topological order in a correlated Chern insulator''}

\author{Joseph Maciejko}
\affiliation{Princeton Center for Theoretical Science, Princeton University, Princeton, New Jersey 08544, USA}

\author{Andreas R\"{u}egg}
\affiliation{Department of Physics, University of California, Berkeley, California 94720, USA}

\date\today

\maketitle

In this supplemental material, we give further details concerning the $\mathbb{Z}_2$ gauge theory description of the interacting Chern insulator [Eq.~(4) of the main text], the $K$-matrix description of the intrinsic topological order in the CI* phase, as well as the numerical solution of the slave-particle Hamiltonian for the Haldane-Hubbard model (an example of interacting Chern insulator) in the mean-field approximation.

\section{Gauge theory representation and saddle-point approximation}

While a detailed study of the path integral representation of the $\mathbb{Z}_2$ slave-spin theory will appear in a separate publication,\cite{z2longpaperSM} here we describe the saddle-point approximation which leads to the final form of the $\mathbb{Z}_2$ lattice gauge theory in Eq.~(4) of the main text. Following the method described in the Appendix A of Ref.~\onlinecite{senthil2000}, one can derive an exact path integral representation of the theory defined by the Hamiltonian (3) of the main text. In order to fully define a theory however, we need to specify not only the Hamiltonian but also the Hilbert space.\cite{WenBook} The Hamiltonian (3) is written in terms of slave-fermion $f_{r\alpha},f_{r\alpha}^\dag$ and slave-spin $\tau^x,\tau^z$ operators which act on a many-body Hilbert space (or Fock space) formed by the tensor product $\mathcal{H}_f\otimes\mathcal{H}_\tau$ of the slave-fermion many-body Hilbert space $\mathcal{H}_f$ and the slave-spin many-body Hilbert space $\mathcal{H}_\tau$. However, this tensor product Hilbert space is larger than the original electron many-body Hilbert space $\mathcal{H}_e$. Therefore, we really need to consider a theory defined by the Hamiltonian (3) acting on a subset $\mathcal{H}_e\cong\mathcal{H}_\textrm{phys}\subset
\mathcal{H}_f\otimes\mathcal{H}_\tau$ of the tensor product Hilbert space. This is the physical Hilbert space
$\mathcal{H}_\textrm{phys}=\{|\psi_\textrm{phys}\rangle\}$ or space of physical states $|\psi_\textrm{phys}\rangle$. The physical states must satisfy the constraint that if there are zero or two slave-fermions $\sum_\alpha f_{r\alpha}^\dag f_{r\alpha}=0,2$ on site $r$, then the Ising variable $\tau^z_r=1$, whereas if there is a single slave-fermion $\sum_\alpha f_{r\alpha}^\dag f_{r\alpha}=1$ on site $r$, then the Ising variable $\tau^z_r=-1$. In other words, the slave-fermions and slave-spins do not fluctuate independently of each other but are tied together by a local constraint. In a path integral representation, this constraint is most effectively implemented by a projector operator $P$ such that $P|\psi_\textrm{phys}\rangle=|\psi_\textrm{phys}\rangle$ and $P^2=1$. In operator form, the projector can be written as
\begin{align}
P=\prod_r\frac{1}{2}\left[1+(-1)^{\sum_\alpha f_{r\alpha}^\dag f_{r\alpha}+\frac{1}{2}(\tau^z_r-1)}\right].
\label{eq:P}
\end{align}
When computing the partition function $Z$, we should sum only over physical states, which is implemented by defining the partition function as
\begin{align}
Z=\Tr\left(e^{-\beta H}P\right),
\end{align}
where $\beta$ is the inverse temperature. One can then use the Suzuki-Trotter formula and the fact that the Hamiltonian $H$ commutes with the projector $P$ (as can be checked explicitly) to derive a path integral representation for $Z$,
\begin{align}\label{Zpathintegral}
Z=\int\prod_{\tau=1}^M\prod_{r\alpha}d\bar{f}_{r\alpha\tau} df_{r\alpha\tau}\sum_{\{\tau^x_{r\tau}\}}
\sum_{\{\tau^z_{r\tau}\}}\sum_{\{\sigma_{r\tau}\}}e^{-S},
\end{align}
with the imaginary time action $S=S_\tau^f+S_\tau^\textrm{Ising}
+\epsilon\sum_{\tau=1}^MH(\tau^x_\tau,\tau^z_\tau,\bar{f}_\tau,f_\tau)$ where
\begin{align}
S_\tau^f&=\sum_{\tau=1}^M\sum_{r\alpha}
\bar{f}_{r\alpha\tau}(\sigma_{r,\tau+1}f_{r\alpha,\tau+1}
-f_{r\alpha\tau}),\\
S_\tau^\textrm{Ising}&=-\frac{i\pi}{4}\sum_{\tau=1}^M
\sum_r(1-\tau^z_{r\tau})[\tau^x_{r\tau}-\tau^x_{r,\tau-1}
-(1-\sigma_{r\tau})],
\end{align}
with the boundary conditions $f_{r\alpha,\tau=M+1}=-f_{r\alpha,\tau=1}$, $\tau^x_{r,\tau=M}=\tau^x_{r,\tau=0}$, and $\sigma_{r,\tau=M+1}=\sigma_{r,\tau=1}$, where $\sigma_{r\tau}$ is a new Ising variable whose role is to implement the local constraint between slave-fermions and slave-spins. We define $\epsilon=\beta/M$ as a short-time cutoff where $M\gg 1$ is the number of imaginary time slices (indexed by $\tau=1,2,\ldots,M$) in the Suzuki-Trotter expansion. The four-operator term $\sim\tau^x\tau^x f^\dag f$ in the Hamiltonian $H$ appearing in Eq.~(\ref{Zpathintegral}) can then be decomposed via the Hubbard-Stratonovich procedure into a sum of two-operator terms $\propto\tau^x\tau^x$ and $\propto f^\dag f$ coupled to an auxiliary field $\chi$. The partition function then takes the form
\begin{align}\label{PathHS}
Z=\int\prod_{\tau=1}^M\prod_{r\alpha}d\bar{f}_{r\alpha\tau} df_{r\alpha\tau}
\int\prod_{rr'}d\chi_{rr'}
\sum_{\{\tau^x_{r\tau}\}}
\sum_{\{\tau^z_{r\tau}\}}\sum_{\{\sigma_{r\tau}\}}e^{-S},
\end{align}
where the action is
\begin{align}\label{Action}
S=S_\tau^f+S_\tau^\textrm{Ising}+\frac{\epsilon U}{4}
\sum_{\tau=1}^M\sum_r(\tau^z_{r\tau}+1)
-\epsilon\sum_{\tau=1}^M\sum_{rr'}
\left(\frac{1}{t}\chi_{rr'}^{(1)}(\tau)\chi_{rr'}^{(2)}(\tau)
+\chi_{rr'}^{(1)}(\tau)\tau^x_{r\tau}\tau^x_{r'\tau}
+\chi_{rr'}^{(2)}\sum_\alpha\Gamma_{rr'}
\bar{f}_{r\alpha\tau}f_{r'\alpha\tau}\right),
\end{align}
where we wrote $t_{rr'}\equiv t\Gamma_{rr'}$ with $t>0$ a positive amplitude, and $\chi_{rr'}^{(1)}$ and $\chi_{rr'}^{(2)}$ are the real scalar auxiliary fields used in the Hubbard-Stratonovich decomposition.

So far all the manipulations have been exact.\cite{senthil2000} In order to obtain the final form of the partition function [Eq.~(4) in the main text] as a $\mathbb{Z}_2$ gauge theory with fermionic and bosonic matter, we need to perform a saddle-point approximation to the path integral over the auxiliary field $\chi_{rr'}$ in Eq.~(\ref{PathHS}). (The idea is described in the paragraphs between Eq.~(54) and Eq.~(56) of Ref.~\onlinecite{senthil2000}.) Essentially, if we were to integrate out the slave-fermions and slave-spins in Eq.~(\ref{PathHS}) we would obtain an effective bosonic action of the Ginzburg-Landau type for the scalar field $\chi_{rr'}$. We could then find the classical minima of this Ginzburg-Landau action, and evaluate the path integral approximately by considering small fluctuations about the classical minimum of least action. While this program is difficult to carry out in practice, the simplest saddle-point which respects the translational symmetry of the problem corresponds to the uniform ansatz $\langle\chi_{rr'}^{(1)}\rangle=\chi_1$, $\langle\chi_{rr'}^{(2)}\rangle=\chi_2$ where $\chi_1$ and $\chi_2$ are two real numbers. However, this saddle-point violates the local (gauge) $\mathbb{Z}_2$ symmetry  of the original Hamiltonian [Eq.~(3) of the main text] under $\tau^x_r\rightarrow\varepsilon_r\tau^x_r$, $f_{r\alpha}\rightarrow\varepsilon_r f_{r\alpha}$, $f_{r\alpha}^\dag\rightarrow\varepsilon_r f_{r\alpha}^\dag$ where $\varepsilon_r=\pm 1$ parameterizes a position-dependent $\mathbb{Z}_2$ gauge transformation. Breaking this $\mathbb{Z}_2$ gauge symmetry is not allowed, because it would correspond to going outside the physical Hilbert space $\mathcal{H}_\textrm{phys}$. In order to preserve the $\mathbb{Z}_2$ gauge symmetry while preserving the simplicity of the uniform saddle-point, we keep the \emph{magnitude} of $\chi_{rr'}$ constant while allowing for \emph{sign} fluctuations:
\begin{align}
\chi_{rr'}^{(1)}(\tau)=\chi_1\sigma_{rr'}(\tau),\hspace{5mm}
\chi_{rr'}^{(2)}(\tau)=\chi_2\sigma_{rr'}(\tau),
\end{align}
where $\sigma_{rr'}(\tau)=\pm 1$ is a dynamical Ising variable living on the spatial links of the lattice which corresponds to a $\mathbb{Z}_2$ gauge field. Under the gauge transformation described above, the $\mathbb{Z}_2$ gauge field transforms as $\sigma_{rr'}\rightarrow\varepsilon_r
\sigma_{rr'}\varepsilon_{r'}$, which ensures that the action (\ref{Action}) is gauge invariant. The term $\sim\frac{1}{t}\chi_{rr'}^{(1)}\chi_{rr'}^{(2)}$ in Eq.~(\ref{Action}) is a constant and can be dropped. Finally, the spatial $\mathbb{Z}_2$ gauge field $\sigma_{rr'}$ can be combined with $\sigma_{r\tau}$ to form a full-fledged spacetime $\mathbb{Z}_2$ gauge field $\sigma_{ij}$ where $i,j$ are spacetime indices, and we obtain the $\mathbb{Z}_2$ gauge theory in 3D Euclidean spacetime of Eq.~(4) in the main text.

\section{$K$-matrix description of topological order}

The CI* phase corresponds to the deconfined phase of the $\mathbb{Z}_2$ gauge theory in Eq.~(4) of the main text. As mentioned in the main text, in the deconfined phase we can neglect\cite{senthil2000} the Berry phase term as far as low-energy, long-wavelength properties are concerned. The first step in deriving a low-energy effective theory for the CI* phase is to integrate out the gapped slave-spins $\tau^x$, which generates a Maxwell term for the $\mathbb{Z}_2$ gauge field. In the $U(1)$ description, by symmetry and power counting we obtain the continuum Lagrangian
\begin{align}\label{ContinuumL}
\mathcal{L}=-\frac{1}{2g^2}f_{\mu\nu}f^{\mu\nu}+\frac{p}{2\pi}\epsilon^{\mu\nu\lambda}b_\mu\partial_\nu a_\lambda+\mathcal{L}_f\left(\bar{f}_\alpha,(\partial_\mu+ia_\mu+ieA_\mu)f_\alpha\right),
\end{align}
where $f_{\mu\nu}=\partial_\mu a_\nu-\partial_\nu a_\mu$ is the field strength for the internal compact gauge field $a_\mu$, $g$ is the internal gauge coupling (which is a function of $\kappa$), and $p=2$. The fermionic action $\int d^3x\,\mathcal{L}_f$ corresponds to the continuum limit of the lattice action $S_f$ in the main text. Because the $BF$ term in Eq.~(\ref{ContinuumL}) is more relevant than the Maxwell term $f_{\mu\nu}f^{\mu\nu}$, we can ignore the Maxwell term in the long-wavelength limit. The only feature of $\mathcal{L}_f$ which is important for the universal, topological properties of the CI* phase is that it describes a fermionic system with total Chern number 2, i.e., with total Hall conductance 2 in units of $e^2/h$ with $e$ the electron charge and $h$ Planck's constant (in the following we set $\hbar=1$). In other words, the total electromagnetic current $J^\mu$ should obey the linear response equation
\begin{align}\label{HallResponse}
J^\mu=\frac{\delta S_f}{\delta A_\mu}
=\frac{2e^2}{2\pi}\epsilon^{\mu\nu\lambda}\partial_\nu A_\lambda.
\end{align}
In the hydrodynamic approach,\cite{wen1995,lu2012} this property is described by introducing two conserved $U(1)$ currents
\begin{align}\label{currents}
j_\uparrow^\mu=\frac{1}{2\pi}\epsilon^{\mu\nu\lambda}\partial_\nu a^\uparrow_\lambda,\hspace{5mm}
j_\downarrow^\mu=\frac{1}{2\pi}\epsilon^{\mu\nu\lambda}\partial_\nu a^\downarrow_\lambda,
\end{align}
such that $J^\mu=e(j_\uparrow^\mu+j_\downarrow^\mu)$ and $a^\uparrow_\mu$ and $a^\downarrow_\mu$ are hydrodynamic or internal gauge fields. As explained for example in Ref.~\onlinecite{lu2012}, it is necessary to introduce two currents for a fermionic system with Chern number 2, otherwise the fermionic statistics of the microscopic constituents is not correctly accounted for. An effective Lagrangian which produces Eq.~(\ref{HallResponse}) is
\begin{align}\label{TopLagrangian}
\mathcal{L}=\frac{p}{2\pi}\epsilon^{\mu\nu\lambda}b_\mu\partial_\nu a_\lambda+\frac{1}{4\pi}\epsilon^{\mu\nu\lambda}
a^\uparrow_\mu\partial_\mu a^\uparrow_\lambda
+\frac{1}{4\pi}\epsilon^{\mu\nu\lambda}
a^\downarrow_\mu\partial_\mu a^\downarrow_\lambda
+\frac{1}{2\pi}\epsilon^{\mu\nu\lambda}(a_\mu-eA_\mu)\partial_\nu(a^\uparrow_\lambda
+a^\downarrow_\lambda).
\end{align}
Indeed, the equations of motion obtained by varying the Lagrangian (\ref{TopLagrangian}) with respect to the fields $a_\mu,b_\mu,a_\mu^\uparrow,a_\mu^\downarrow$ are
\begin{align}
0&=\frac{p}{2\pi}\epsilon^{\mu\nu\lambda}\partial_\nu b_\lambda+\frac{1}{2\pi}\epsilon^{\mu\nu\lambda}\partial_\nu
(a^\uparrow_\lambda+a^\downarrow_\lambda),\\
0&=\frac{p}{2\pi}\epsilon^{\mu\nu\lambda}\partial_\nu a_\lambda,\\
0&=\frac{1}{2\pi}\epsilon^{\mu\nu\lambda}\partial_\nu
a^\uparrow_\lambda+\frac{1}{2\pi}\epsilon^{\mu\nu\lambda}\partial_\nu(a_\lambda-
eA_\lambda),\\
0&=\frac{1}{2\pi}\epsilon^{\mu\nu\lambda}\partial_\nu
a^\downarrow_\lambda+\frac{1}{2\pi}\epsilon^{\mu\nu\lambda}\partial_\nu(a_\lambda-
eA_\lambda),
\end{align}
and upon using the definition (\ref{currents}) of the currents we obtain the expected Hall response (\ref{HallResponse}). The Lagrangian (\ref{TopLagrangian}) can be put in the standard $K$-matrix form\cite{WenBook}
\begin{align}
\mathcal{L}=\frac{1}{4\pi}\epsilon^{\mu\nu\lambda}K_{IJ}a_\mu^I\partial_\nu a_\lambda^J-\frac{e}{2\pi}\epsilon^{\mu\nu\lambda}
t_I A_\mu\partial_\nu a_\lambda^I,
\end{align}
where $I,J=1,\ldots,4$ with $a_\mu^I=(a_\mu,b_\mu,a_\mu^\uparrow,
a_\mu^\downarrow)$, and the $K$-matrix and electromagnetic charge vector $t$ are (with $p=2$)
\begin{align}
K=\left(\begin{array}{cccc}
0 & 2 & 1 & 1\\
2 & 0 & 0 & 0\\
1 & 0 & 1 & 0\\
1 & 0 & 0 & 1
\end{array}\right),\hspace{5mm}
t=\left(\begin{array}{c}
0\\
0\\
1\\
1
\end{array}\right).
\end{align}
One can always transform $K$ and $t$ to an equivalent representation $K'=W^TKW$ and $t'=W^Tt$ by a $GL(4,\mathbb{Z})$ transformation $W$. In general $K'$ and $t'$ will not be diagonal, but here it is possible to find a $GL(4,\mathbb{Z})$ matrix $W$ given by
\begin{align}
W=\left(\begin{array}{cccc}
1 & -1 & 0 & 0\\
1 & 0 & 0 & 0\\
-1 & 1 & 1 & 0\\
-1 & 1 & 0 & 1
\end{array}\right),
\end{align}
such that $K'$ is diagonal,
\begin{align}
K'=\left(\begin{array}{cccc}
2 & 0 & 0 & 0\\
0 & -2 & 0 & 0\\
0 & 0 & 1 & 0\\
0 & 0 & 0 & 1
\end{array}\right),\hspace{5mm}
t'=\left(\begin{array}{c}
-2\\
2\\
1\\
1
\end{array}\right),
\end{align}
as given in Eq.~(9) of the main text.

\section{Numerical study of the Haldane-Hubbard model}
The Haldane-Hubbard model studied in the second part of the main text is defined on the honeycomb lattice and given by
\begin{equation}
H=-t\sum_{\langle rr'\rangle}\sum_\alpha\left(c_{r\alpha}^{\dag}c_{r'\alpha}+{\rm H.c.}\right)+t_2\sum_{\langle\!\langle rr'\rangle\!\rangle}\sum_\alpha\left(i\nu_{rr'}c_{r\alpha}^{\dag}c_{r'\alpha}+
{\rm H.c.}\right)+U\sum_r n_{r\uparrow}n_{r\downarrow}.
\label{eq:H_Haldane_Hubbard}
\end{equation}
Here, following Ref.~\onlinecite{haldane1988}, $\nu_{rr'}=\pm 1$ assigns phase factors $\pm i$ to the second-neighbor hopping resulting in a staggered flux pattern which preserves both the original unit cell as well as the six-fold rotation symmetry around the center of a hexagon. We consider half-filling where both spin-up and spin-down electrons fill a Bloch band with the same Chern number $C_{\uparrow}=C_{\downarrow}=\pm 1$. In the slave spin-representation, Eq.~\eqref{eq:H_Haldane_Hubbard} at half-filling takes the form (up to an additive constant)
\begin{equation}
H'=-t\sum_{\langle rr'\rangle}\sum_\alpha\left(\tau_r^x\tau_{r'}^xf_{r\alpha}^{\dag}
f_{r'\alpha}+{\rm H.c.}\right)+t_2\sum_{\langle\!\langle rr'\rangle\!\rangle}\sum_\alpha
\left(i\nu_{rr'}\tau_r^x\tau_{r'}^xf_{r\alpha}^{\dag}
f_{r'\alpha}+{\rm H.c.}\right)+\frac{U}{2}\sum_r\tau_r^z.
\end{equation}
In the physical subspace defined by the projector Eq.~\eqref{eq:P}, $H$ and $H'$ are equivalent. In the mean-field approximation, slave-spins and slave-fermions are decoupled. This results in the following mean-field problem,
\begin{eqnarray}
H_f&=&-t\sum_{\langle rr'\rangle}\sum_\alpha g_{rr'}\left(f_{r\alpha}^{\dag}f_{r'\alpha}+{\rm H.c.}\right)+t_2\sum_{\langle\!\langle rr'\rangle\!\rangle}
\sum_\alpha g_{rr'}'\left(i\nu_{rr'}f_{r\alpha}^{\dag}f_{r'\alpha}+{\rm H.c.}\right),\\
H_{\tau}&=&\sum_{\langle rr'\rangle}J_{rr'}\tau_r^x\tau_{r'}^x+\sum_{\langle\!\langle rr'\rangle\!\rangle}J_{rr'}'
\tau_r^x\tau_{r'}^x+\frac{U}{2}\sum_r\tau_r^z,
\label{eq:Htau}
\end{eqnarray}
with self-consistency conditions
\begin{eqnarray}
g_{rr'}&=&\langle\tau_r^x\tau_{r'}^x\rangle,\mbox{ $(r,r')$ nearest neighbors;}\label{eq:g}\\
g_{rr'}'&=&\langle\tau_r^x\tau_{r'}^x\rangle,\mbox{ $(r,r')$ second neighbors;}\label{eq:gp}\\
J_{rr'}&=&-t\sum_{\alpha}\left(\langle f_{r\alpha}^{\dag}f_{r'\alpha} \rangle+{\rm c.c.}\right),\mbox{ $(r,r')$ nearest neighbors;}\\
J_{rr'}'&=&t_2\sum_{\alpha}\left(i\nu_{rr'}\langle f_{r\alpha}^{\dag}f_{r'\alpha} \rangle+{\rm c.c.}\right),\mbox{ $(r,r')$ second neighbors.}
\end{eqnarray}
To obtain the slave-spin correlation functions Eqs.~\eqref{eq:g} and \eqref{eq:gp}, we used the semi-classical approximation to the transverse-field Ising model Eq.~\eqref{eq:Htau},\cite{ruegg2010} generalized to inhomogeneous mean-field solutions.\cite{ruegg2012} For large $t_2\sim t$ and intermediate to large interaction $U$, one indeed finds a mean-field solution which corresponds to the CI* phase.

\subsection{Topological ground-state degeneracy and spin-charge separation in the CI*}
The numerical analysis of the topological ground-state degeneracy and the spin-charge separation shown in Fig.~1 of the main text follows Ref.~\onlinecite{ruegg2012}. On the torus, four different ground-states are obtained by simultaneously considering periodic or anti-periodic boundary conditions for the slave-fermions and slave-spins (the boundary conditions of the original electrons are always periodic). This corresponds to placing or removing a vison ($\mathbb{Z}_2$ flux) through the two non-contractible loops of the torus. The presence of such a vison is indicated by $\nu_l=-1$, ($l=1,2$) where
\begin{equation}
\nu_l=\prod_{\mbox{\scriptsize along }{\boldsymbol a}_l}{\rm sign}(g_{rr'}),
\end{equation}
and the product extends around a loop along the unit cell vector $\boldsymbol{a}_l$.
For a finite system, there is an exponentially small splitting of the energy among these four states, as shown in Fig.~1(a) of the main text.

To study the spin-charge separation, we doped a single hole into the system. One can then stabilize mean-field solutions with a vison pair, indicated by $\prod_{\hexagon}{\rm sign}(g_{rr'})=-1$ for two well-separated hexagons. The slave-fermions see the vison as an external $\pi$ flux which (because of the non-vanishing Chern number) induces bound states. For sufficiently well-separated visons, one can choose a basis in which all the charge of the doped hole is localized at the position of the first vison while the spin is localized at the position of the second vison, see Fig.~1(b) of the main text.

\begin{figure}
\includegraphics[width=0.45\linewidth]{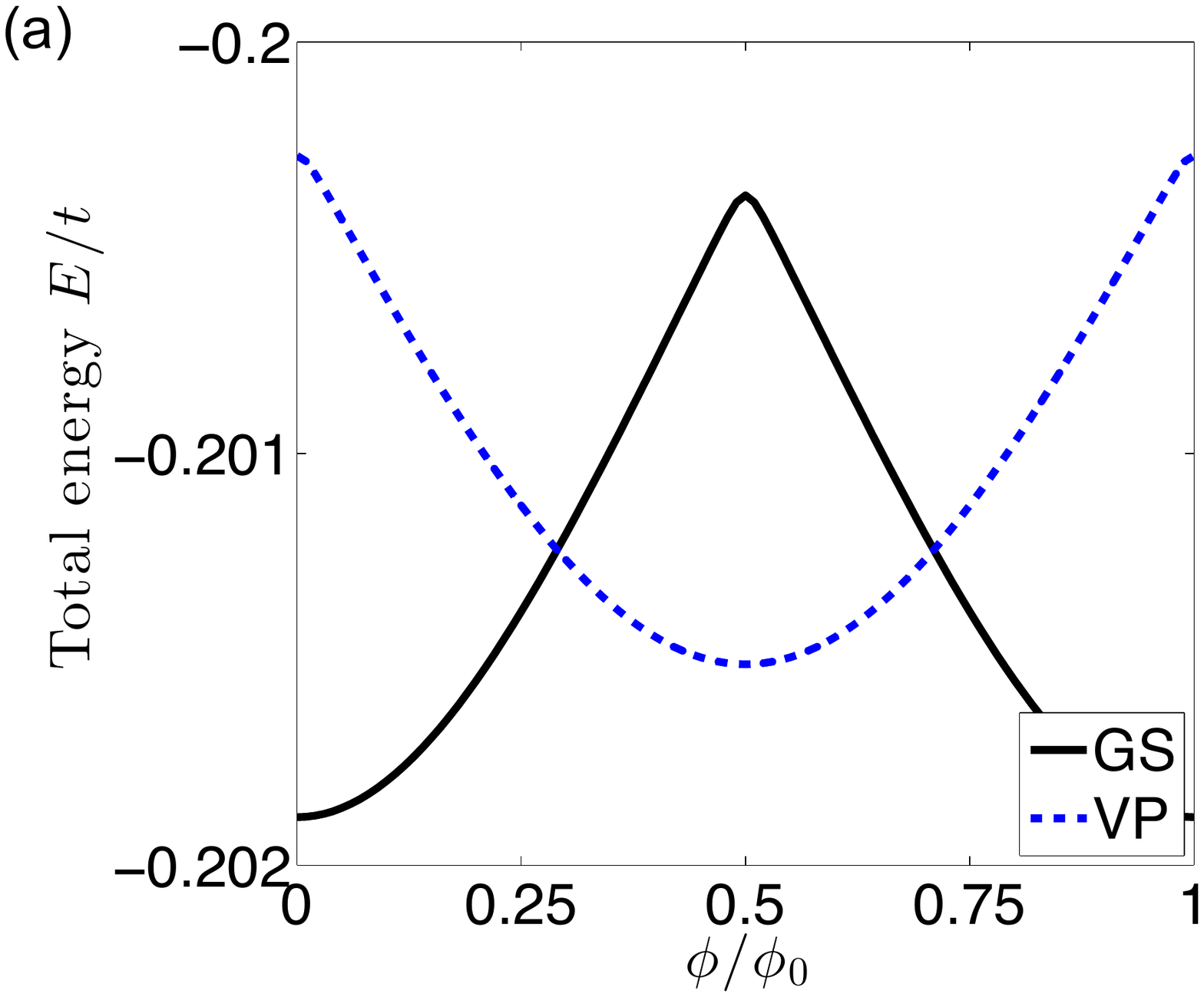}
\includegraphics[width=0.43\linewidth]{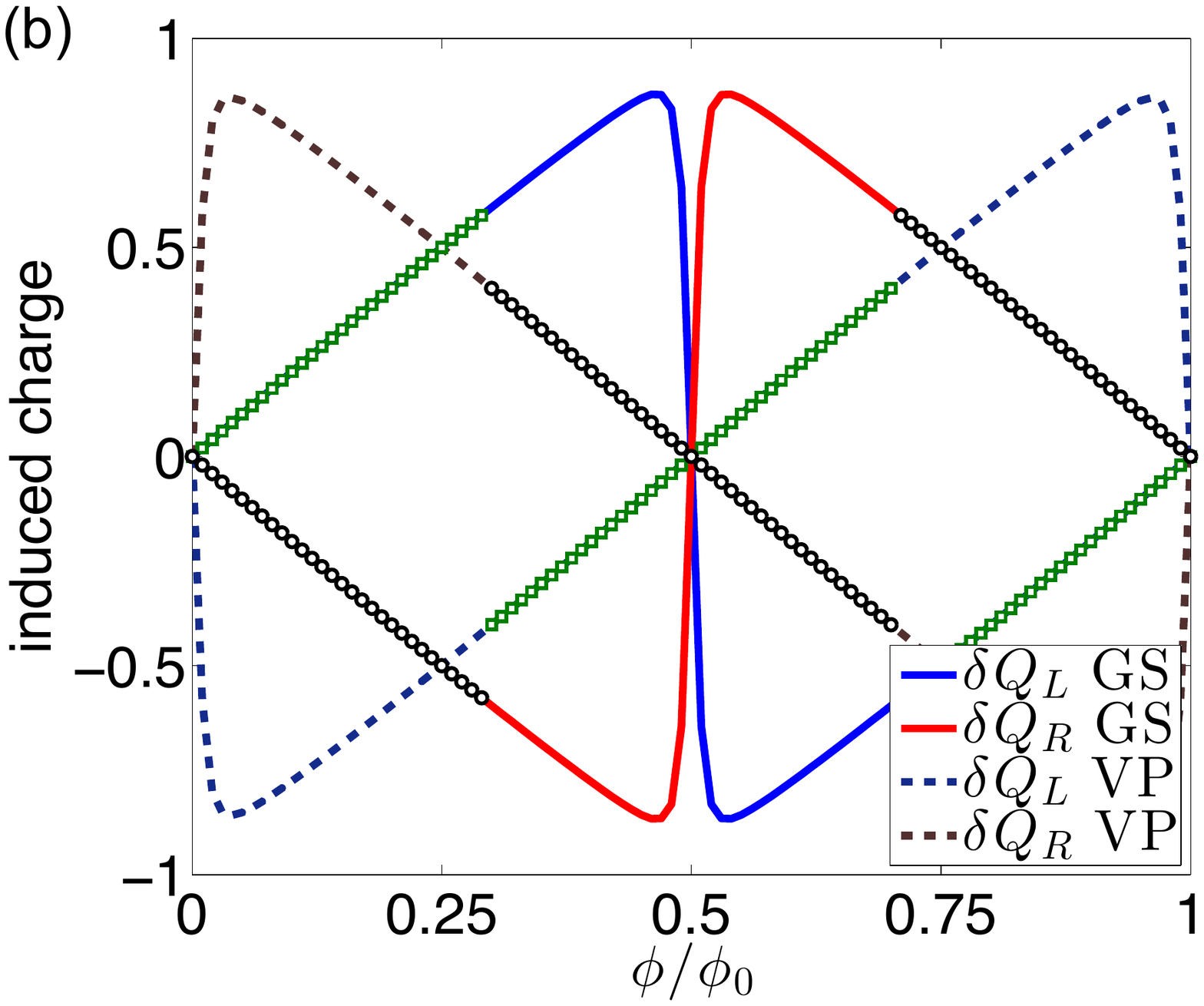}
\caption{(a) Total energy as function of external flux $\phi/\phi_0$ in the charge pumping setup described in the text. Shown are the energy of a solution without visons (GS) and the energy of a solution with a vison pair at the location of the external fluxes (VP). (b) Induced charge at the location of the external fluxes (L) and (R) for the two solutions (GS) and (VP).}
\label{fig:CP}
\end{figure}

\subsection{Charge pumping in the CI*}
To study the charge pumping in the CI* phase, we considered a finite system with $16\times 16$ unit cells and periodic boundary conditions and used $t_2=t$ and $U=26t$. A flux $\phi$ was threaded through the center of one hexagon while a flux $-\phi$ was threaded through the center of another hexagon separated by 8 unit cells from the first one.

We compared two different types of (meta-) stable solutions. In the first solution, the $\mathbb{Z}_2$ gauge field remains in its ground state configuration, i.e., $\prod_{\hexagon}{\rm sign}(g_{rr'})=1$ around each elementary hexagon. We denote this solution as GS in Fig.~\ref{fig:CP}. The second solution places a vison pair at the locations of the external fluxes, i.e., the product $\prod_{\hexagon}{\rm sign}(g_{rr'})=-1$ for the two hexagons pierced by the external fluxes. This solution is denoted as VP in Fig.~\ref{fig:CP}. 

As demonstrated in Fig.~\ref{fig:CP}(a), the vison pair solution is lowest in energy for a range of external flux $1/4\lesssim\phi/\phi_0\lesssim 3/4$. In this parameter regime, it is energetically favorable to (partially) screen the external fluxes by spontaneously creating a vison pair. Note that within the mean-field approximation, there is a level crossing between the two solutions GS and VP. We expect that if the dynamics of the $\mathbb{Z}_2$ gauge field is taken into account more accurately, the level crossing will turn into an avoided level crossing. Figure~\ref{fig:CP}(b) shows the induced charge at the positions of the two external fluxes (L) and (R) for the two solutions GS and VP. It becomes clear that the two solutions VP and GS are shifted by $\phi_0/2$ with respect to each other.

\bibliography{qah}